\documentclass[aps,preprint,tightenlines,superscriptaddress,showpacs]{revtex4}
\usepackage{graphicx}
\usepackage{epsfig}
\usepackage{dcolumn}
\usepackage{bm}
\usepackage{color}
\usepackage{overpic}
\usepackage{times}
\usepackage{amsmath}
\usepackage{tabularx}
\usepackage{multirow}
\usepackage{longtable}
\usepackage{textcomp}
\usepackage[colorlinks,linkcolor=red,anchorcolor=blue,citecolor=blue]{hyperref}
\usepackage{comment}

\usepackage{ulem}

\graphicspath{{./}}

\newcommand{\EE}{e^+e^-}

\newcommand{\bbb}{B\overline{B}}

\begin{document}

\title{\boldmath A novel method of reconstructing semileptonic $B$ decays}

\author{Long-Ke~Li}
 \email{lilongke@ihep.ac.cn}
\affiliation{Institute of High Energy Physics, Chinese Academy of
Sciences, 19B Yuquan Road, Beijing 100049, China}

\author{Chang-Zheng~Yuan}
 \email{yuancz@ihep.ac.cn}
\affiliation{Institute of High Energy Physics, Chinese Academy of
Sciences, 19B Yuquan Road, Beijing 100049, China}
\affiliation{University of Chinese Academy of Sciences, 19A Yuquan
Road, Beijing 100049, China}

\begin{abstract}

Semileptonic decays of charged and neutral $B$ mesons play a
critical role in the determination of the magnitudes of the
CKM-matrix elements $V_{cb}$ and $V_{ub}$, and in the test of the
lepton universality which is a basic assumption of the Standard
Model. Due to the missing neutrino in the semileptonic decays, the
measurements depend strongly on the tag efficiency of the $B$
meson in the recoil side which is at a few per mille to a few per
cent level, and this values a limiting factor for high precision measurement
and high sensitivity test mentioned above. 
We develop a new method
of reconstructing semileptonic decays of the $B$ mesons by
introducing the $B$ momentum information calculated from the $B$
decay vertex and the interaction point, as well as the kinematic
information of $\EE\to \bbb$ at the $B$-factories such as BaBar,
Belle, and Belle II. As this method does not depend on the
reconstruction of the $B$ meson in the recoil side, the gain in
the efficiency could be as large as two orders of magnitudes. This
makes high precision measurement of CKM-matrix elements $|V_{cb}|$
and $|V_{ub}|$ and high precision test of lepton universality at
$B$-factories more promising. We present the algorithms for the
semileptonic decays of neutral and charged $B$ mesons separately,
as they are affected by the magnetic field in the detector
differently.

\end{abstract}


\maketitle

\section{Introduction}

Semileptonic (SL) decays of charged and neutral $B$ mesons,
proceed via leading-order weak interactions, play a critical role
in the determination of the magnitudes of the
Cabibbo-Kobayashi-Maskawa (CKM) matrix~\cite{bib:CKM} elements
$V_{cb}$ and $V_{ub}$, which impact most studies of flavor
physics and $CP$-violation in the quark sector, along with an understanding of properties
of the $b$ quark bound in a meson. And it also is very important in the test of the lepton
universality which is the basic assumption of the Standard Model. The leptonic part in the effective
Hamiltonian and the decay matrix element factorizes from hadronic
part, and QCD corrections can only occur in the $b\to q$
current~\cite{bib:PBF}, as an important feature of SL $B$ decays.

$B$-factories collected a large $B$ meson sample to study $B$
physics. For example, there are 772 million $B\bar{B}$ pairs at
the Belle experiment and 471 million $B\bar{B}$ pairs at the BaBar
experiment~\cite{bib:PBF}. However, in studying $B$ decays with an
undetected neutrino or a missing particle (such as a neutron or
$K_L$ meson), the $B$ meson in the recoil side must be fully or
partially reconstructed in order to infer the information of the
missing particle. This $B$-tag technique is not very efficient,
because all $B$ meson ($b$ quark) weak decays are Cabibbo
suppressed and there are no dominant decay modes can be used for
the reconstruction.

The method of identifying signal candidates of SL $B$ decays using
fully reconstructed $B$ decays in the recoil side has been
employed in exclusive $\bar{B}\to X_{u}\ell^{-} \bar{\nu}_{\ell}$
decays (where $X_{u}$ denotes a light meson containing a $u$
quark, and $\ell$ denotes an electron or muon) by
CLEO~\cite{bib:SLatCLEO1,bib:SLatCLEO2},
BaBar~\cite{bib:SLatBaBar1,bib:SLatBaBar2,bib:SLatBaBar3,bib:SLatBaBar4,bib:SLatBaBar5}
and Belle~\cite{bib:SLatBelle1,bib:SLatBelle2,bib:SLatBelle3}. In
such analyses, the missing energy and momentum of the whole event
are used to reconstruct the neutrino from signal SL decays. In a
partial reconstruction of $B\to D^{(*)}\ell\nu$ decays as the
tagging mode (SL tag), since there are two neutrinos present in
the event, the kinematics cannot be fully
constrained~\cite{bib:partial_tag}.

Belle developed a full reconstruction tool to tag $B$ decays using
a multivariable analysis based on a neural network algorithm named
NeuroBayes~\cite{bib:NeuroBayes}, in which, 1104 exclusive decay-channels 
were reconstructed, amploying 71 neural networks. An
overall efficiency of 0.28\% for $B^{\pm}$ and of 0.18\% for $B^0$
mesons is achieved, which is an improvement by roughly 
a factor of two comparing to the efficiency of the cut-based classical reconstruction algorithm. 
Recently, a new tagging method, full exclusive
interpretation (FEI)~\cite{bib:FEI}, based on machine learning,
has been developed at Belle and Belle II experiments. The FEI
reconstructs more than 100 explicit decay channels, leading to
$\mathcal{O}$(10,000) distinct decay-chains. It achieves the
maximum tagging efficiency to 0.76\%~(1.80\%) for $B^+$ and
0.46\%~(2.04\%) for $B^0$ with hadronic~(SL) tag at
Belle~\cite{bib:BToSLwithFEI}. The $B$ tagging efficiency is at
level of $\mathcal{O}$($10^{-2}$) or even
$\mathcal{O}$($10^{-3}$), which means only a very small fraction
($10^{-3}\sim 10^{-2}$) of the data sample is used in the
measurement of SL decays of the $B$ meson. Considering the
advancement of FEI, it is hard to improve the tagging efficiency
significantly in the further along this line. Therefore, new
method should be developed to improve the efficiency of
reconstructing the SL $B$ decays.

As we know, to measure the $CP$ violation in $B$ decays, the $B$
meson decay and production vertices can be well determined at the
$B$ factories. If there are at least two charged tracks in the
final states of $B$ decays~\cite{footnote1}, $B$ decay vertex,
$\vec{V}_B$=($x_B$, $y_B$, $z_B$), can be determined by a vertex
constraint fit. The interaction point~($IP$),
$\vec{V}_{IP}$=($x_{IP}$, $y_{IP}$, $z_{IP}$), which can be
measured with non-$\bbb$ events, can be regarded as the $B$ meson
production vertex, as the lifetime of $\Upsilon(4S)$, mother of
the $B$ mesons, is very short. At Belle experiment, $IP$ is
time-dependent, and is calculated every 10,000 events to take into
account an observed variation of $IP$ position during data taking.

The $B$ meson decay and production vertices will allow us not only
to calculate the lifetime of the $B$ mesons, but also to provide
additional constraints, besides energy-momentum conservation
constraint, in SL $B$ decays. Based on these additional
constraints, we present a new method of $B$ meson reconstruction
in this article, which allows to fully determine SL $B$
decays (or hadronic $B$ decays with a missing particle) without
tagging the $B$ meson in the other side at an $\EE$ $B$-factories
running at the $\Upsilon(4S)$ energy. We present the
reconstruction algorithms for SL decays of neutral $B$, positive
and negative charged $B$ mesons, separately, as they are affected
by the magnetic fields in the detector in different ways.

\section{The SL decays of neutral $B$ meson}

\begin{figure*}
  \begin{center}
  \includegraphics[width=0.55\textwidth,height=0.38\textwidth]{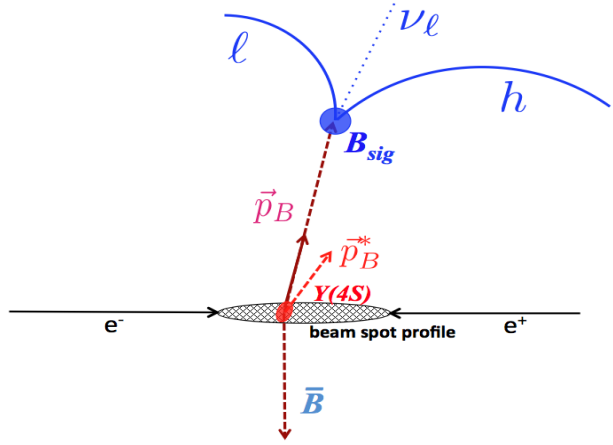}
  \vskip-5pt
\caption{\label{fig:method1}(Color online) A schematic diagram of
SL decays of neutral $B$ meson $B^0_{\rm sig}\to h\ell\nu$, $B$
decay and production vertex information is used to determine the
direction of the $B$ momentum.}
  \end{center}
\end{figure*}

In the case of the SL decays of a neutral $B$, as shown in
Fig.~\ref{fig:method1}, with the magnitude of the $B$ moment
$|\vec{p}_B|$ given, the momentum of the $B^0$ meson is written as
a function of $B$ decay and production vertices:
$\vec{p}_B=|\vec{p}_B| \cdot \frac{\vec{r}}{|\vec{r}|}$, where
momentum unit vector $\frac{\vec{r}}{|\vec{r}|}$ is determined by
$B$ decay vertex and production vertex via
$\vec{r}$=$\vec{V}_{B}-\vec{V}_{IP}$. Thus, energy-momentum
conservation of the SL decays of the $B$ meson gives four
constraints in Eqs.~(\ref{eqn:4C1}-\ref{eqn:4C4}):
\begin{eqnarray}
\sqrt{m_B^2 + |\vec{p}_B|^2 } & = & \sqrt{m_h^2 + |\vec{p}_h|^2} + \sqrt{m_{\ell}^2 + |\vec{p}_{\ell}|^2} +\sqrt{m_{\nu}^2 + | \vec{p}_{\nu} |^2}, \label{eqn:4C1}  \\
|\vec{p}_B| \cdot \frac{r_x}{|\vec{r}|} & = & p_{hx} + p_{\ell x} + p_{\nu x}, \label{eqn:4C2}\\
|\vec{p}_B| \cdot \frac{r_y}{|\vec{r}|} & = & p_{hy} + p_{\ell y} + p_{\nu y}, \label{eqn:4C3}\\
|\vec{p}_B| \cdot \frac{r_z}{|\vec{r}|} & = & p_{hz} + p_{\ell z}
+ p_{\nu z}.  \label{eqn:4C4}
\end{eqnarray}
Therefore this SL decay is fully determined, with known variables
\{$\frac{\vec{p}_B}{|\vec{p}_B|}=\frac{\vec{r}}{|\vec{r}|}$,
$\vec{p}_h$, $\vec{p}_{\ell}$\} and \{$m_B$, $m_h$, $m_{\ell}$,
$m_{\nu}$\} ($m_{\nu}=0$ for neutrino or nominal mass of a missing
particle) and unknown $|\vec{p}_B|$ and $\vec{p}_{\nu}$ (four
variables).

In fact, we have one additional constraint besides those listed in
Eqs.~(\ref{eqn:4C1}-\ref{eqn:4C4}). As the $\bbb$ pair is produced
in $\Upsilon(4S)$ decays, each $B$ meson carries an energy of half
of the $\Upsilon(4S)$ mass in the center-of-mass (CM) frame. Thus,
with known $\Upsilon(4S)$ energy-momentum and the direction of 
the $B$ momentum in the laboratory (LAB)
frame, we can calculate the magnitude of the $B$ momentum, that
is, $|\vec{p}_B|$ in Eqs.~(\ref{eqn:4C1}-\ref{eqn:4C4}). We take
Belle and Belle II cases as example below.

\begin{figure*}
  \begin{center}
  \includegraphics[width=0.4\textwidth]{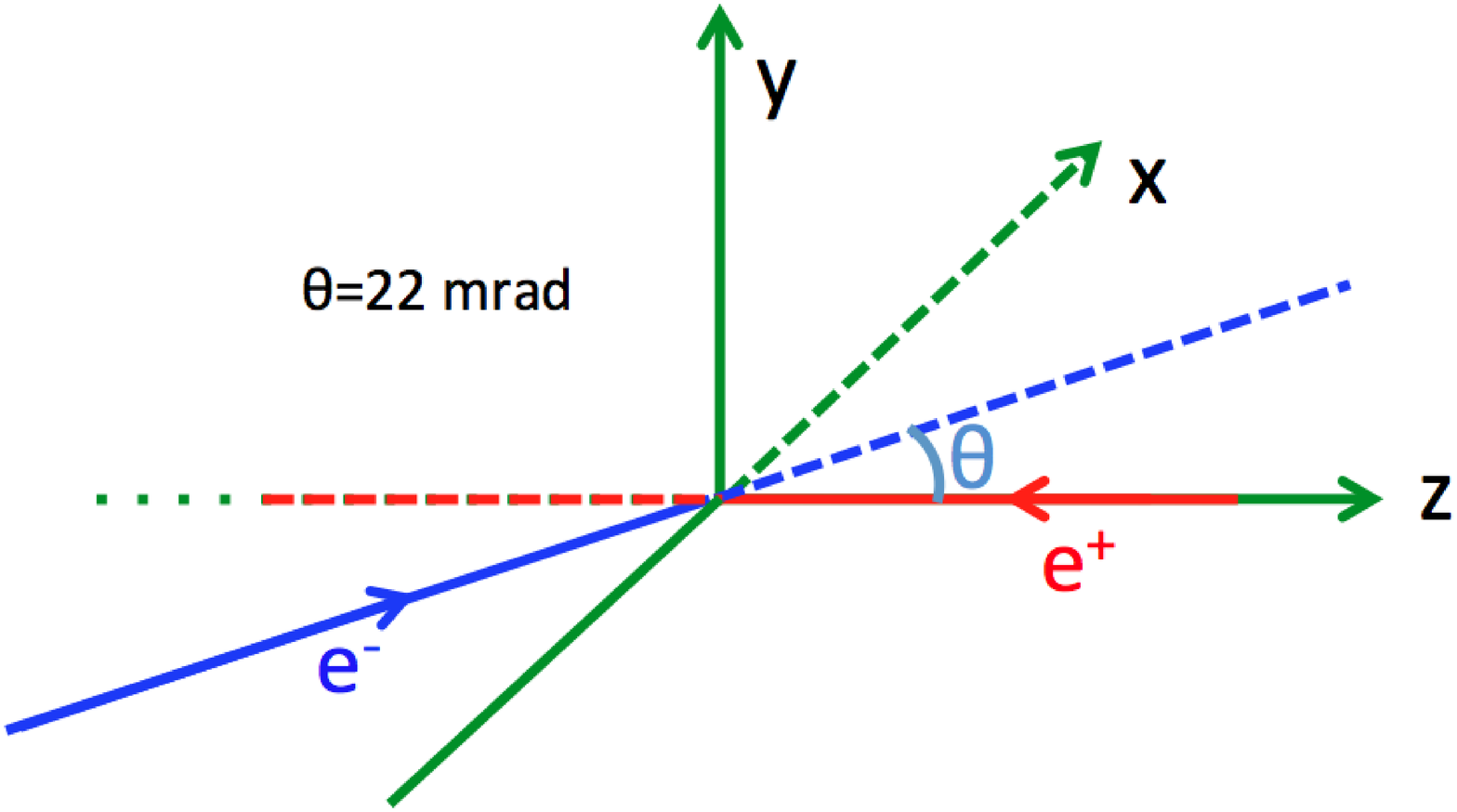}~~~
  \includegraphics[width=0.4\textwidth]{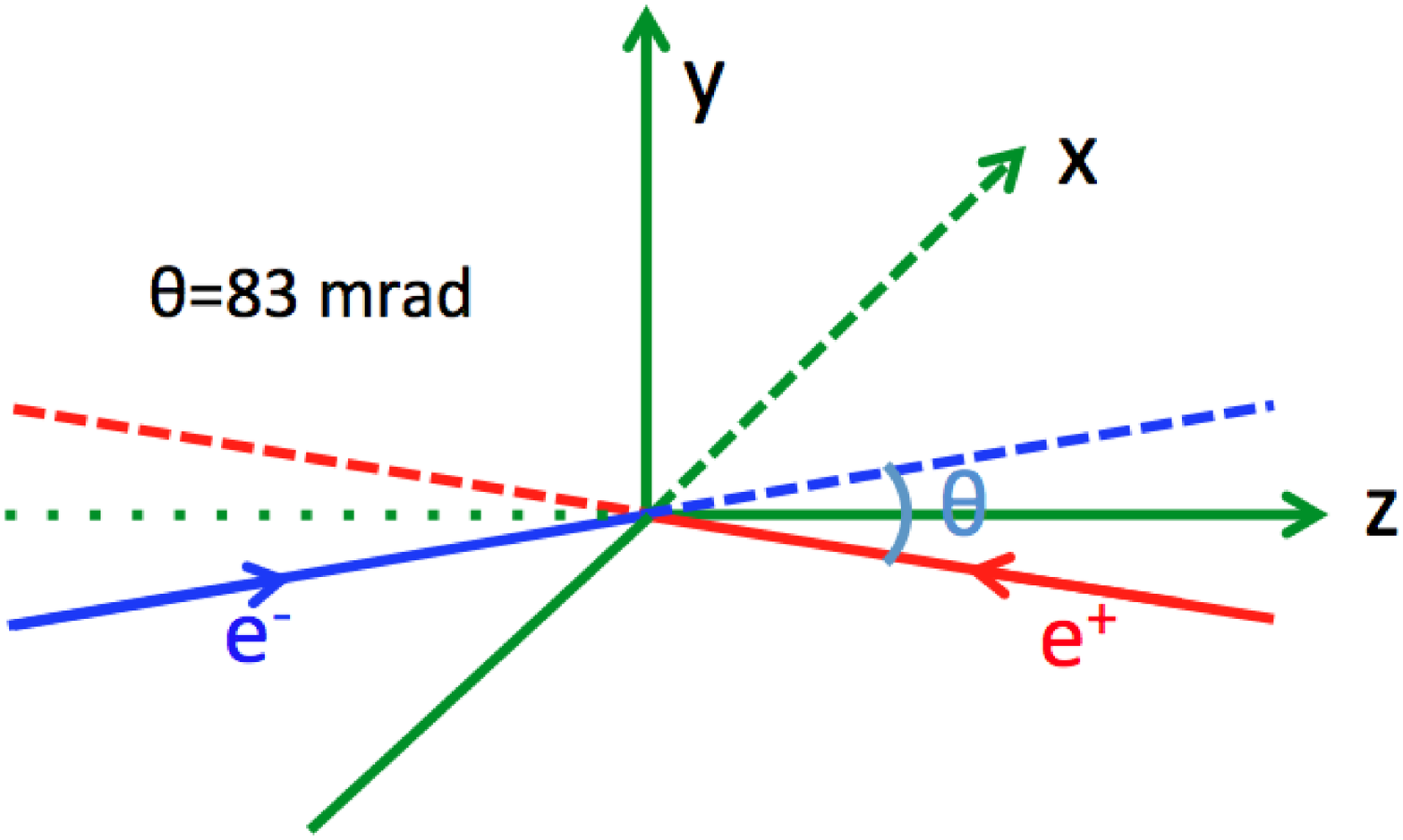}%
\caption{\label{fig:cms}(Color online) Definition of the
laboratory frame at Belle detector (left panel) and Belle II
detector (right panel) with different boost vector $\bm\beta$.}
  \end{center}
\end{figure*}

The $+z$ axis in the laboratory frame, as direction of the nominal
magnetic field $\vec{B}$, are defined differently at Belle and
Belle II detectors, as shown in Fig.~\ref{fig:cms}. The four
momenta of the $\EE$ annihilation CM frame, $P_{\rm cm}$ in
Eqs.~(\ref{eqn:cms1},~\ref{eqn:cms2}), are determined by energy of
electron and positron beams, $E_{-}$=8 (7.004)~GeV and $E_{+}$=3.5
(4.002)~GeV, and their cross angle $\theta$=22 (83)~mrad at Belle
(Belle II) experiment, therefore they both have energy
$\sqrt{s}=\sqrt{2E_{-}E_{+}(1+\cos\theta)}\approx 10.58$~GeV in CM
frame.
\begin{eqnarray}
P_{\rm cm} & = & (E_{-}\sin{\theta}, ~0, ~E_{-}\cos\theta-E_{+}, ~E_{-}+E_{+}) \nonumber \\
& \approx  & (0.176,~ 0,~ 4.498,~ 11.5 )~\text{GeV ~~at Belle,}  \label{eqn:cms1} \\
P_{\rm cm} & = & ((E_{-}+E_{+})\sin{\frac{\theta}{2}}, ~0, ~(E_{-}-E_{+})\cos\frac{\theta}{2}, ~E_{-}+E_{+}) \nonumber \\
& \approx & (0.457,~ 0,~ 2.999,~ 11.006 )~\text{GeV ~~at
Belle~II.}  \label{eqn:cms2}
\end{eqnarray}

The boost vector $\vec{\bm\beta}$ of $B$ meson to mother particle
$\Upsilon(4S)$ CM frame, equals the momentum $-\vec{p}_{\rm cm}$
normalized by total energy as $\vec{\bm\beta}=(\beta_{x},
~\beta_{y}, ~\beta_{z})=\frac{-\vec{p}_{\rm cm}}{E_-+E_+}$. With
four-momentum of the $B$ meson $P$=($\vec{p}_B$, $E$)=($p_x$,
$p_y$, $p_z$, $E$) in LAB frame, the boosted four-momentum
$P^{*}$=($p^{*}_x$, $p^{*}_y$, $p^{*}_z$, $E^*$) in CM frame can
be calculated with a Lorentz boost function.
\begin{eqnarray}
p^{*}_x & = & p_x + \gamma_2 \cdot (\vec{\bm\beta}\cdot\vec{p})\cdot \beta_{x} + \gamma \cdot E \cdot \beta_{x},  \label{eqn:pst_x} \\
p^{*}_y & = & p_y + \gamma_2 \cdot (\vec{\bm\beta}\cdot\vec{p})\cdot \beta_{y} + \gamma \cdot E \cdot \beta_{y}, \label{eqn:pst_y} \\
p^{*}_z & = & p_z + \gamma_2 \cdot (\vec{\bm\beta}\cdot\vec{p})\cdot \beta_{z} + \gamma \cdot E \cdot \beta_{z}, \label{eqn:pst_z} \\
E^{*} & = &  \gamma \cdot( E + \vec{\bm\beta}\cdot\vec{p}).  \label{eqn:Est}
\end{eqnarray}
Here $|\vec{\bm\beta}|^2=\beta_{x}^2+\beta_{y}^2+\beta_{z}^2$ and
$\gamma_2$=$(\gamma -1)/|\vec{\bm\beta}|^2$, and $\gamma$ is
calculated as
\begin{eqnarray}
\gamma & = & \frac{1}{\sqrt{1-|\vec{\bm\beta}|^{2}}} = \frac{E_{-}+E_{+}}{\sqrt{2 E_{-}E_{+}(1+\cos\theta)}}.
\end{eqnarray}
Thus, with these calculations and $E^{*}$=$\sqrt{s}/2$,
Eq.~(\ref{eqn:Est}) becomes
\begin{eqnarray}
\frac{\sqrt{s}}{2\gamma} & = & \sqrt{ |\vec{p}_B|^2 + m_B^2 } +
\dfrac{\beta_x  r_x + \beta_z  r_z}{\sqrt{r_x^2+r_y^2+r_z^2}}
\cdot |\vec{p}_B| . \label{eqn:eqn1}
\end{eqnarray}
Therefore, combining Eq.~(\ref{eqn:eqn1}) with four-momentum
conservation in Eqs.~(\ref{eqn:4C1}-\ref{eqn:4C4}), the SL $B$ decays 
can be fully determined even with a floated $B$ meson
mass. The fractions of signal and background components thus can
be extracted by fitting the candidate $m_B$ distribution.

So far we did not use any information of the $\bar{B}$ decays
recoiling against the $B$ meson in signal side. In fact the above
analysis can be extended to the recoil side by reconstructing
another $B$ vertex and applying the same technique by assuming
there is a missing particle with unknown mass (rather than a zero
mass neutrino in the signal side). As the momentum of $\bar{B}$ in
the CM frame, $\vec{p}^{\ *}_{\bar{B}}$, and $\vec{p}^{\ *}_{B}$
are back-to-back, this constraint can be used to suppress
backgrounds and improve the precision of the $IP$. This additional
constraint is not considered in this article.

To summarize the calculations above, let's label four-momentum of
the final particles as ($p_{ix}$, $p_{iy}$, $p_{iz}$,
$\sqrt{p_i^2+m_i^2}$) (where $i$=$h$, $\ell$, and $\nu$) in LAB
frame. Here $\nu$ can be a neutrino with $m_{\nu}$=$0$ or a
missing particle with $m_{\nu}$ set to its nominal mass. The
initial four-momentum of $B$ meson is ($ |\vec{p}_B|\cdot
r_x/|\vec{r}|$, $|\vec{p}_B|\cdot r_y/|\vec{r}|$,
$|\vec{p}_B|\cdot r_z/|\vec{r}|$, $\sqrt{m^2+|\vec{p}_B|^2}$)
where $\vec{r}/|\vec{r}|$ is the unit vector of the momentum
direction of the reconstructed neutral $B$ meson. We have a
modified form of Eqs.~(\ref{eqn:4C1}-\ref{eqn:4C4},\ref{eqn:eqn1})
as
\begin{eqnarray}
f_{x}: &~~&  p_{\nu x} = |\vec{p}_B| \cdot \frac{r_x}{|\vec{r}|} - (p_{hx} + p_{\ell x}) ,  \label{eqn:v4cfit10}\\
f_{y}: &~~&  p_{\nu y} = |\vec{p}_B| \cdot \frac{r_y}{|\vec{r}|} - (p_{hy} + p_{\ell y}) , \label{eqn:v4cfit20}\\
f_{z}: &~~&  p_{\nu z} = |\vec{p}_B| \cdot \frac{r_z}{|\vec{r}|} - (p_{hz} + p_{\ell z}) , \label{eqn:v4cfit30}\\
f_{e}: &~~&  \sqrt{m_h^2 + |\vec{p}_h|^2} + \sqrt{m_{\ell}^2 + |\vec{p}_{\ell}|^2} + \sqrt{m_{\nu}^2 + |\vec{p}_{\nu}|^2 } =  \frac{\sqrt{s}}{2\gamma} - \frac{\beta_x  r_x+ \beta_z  r_z}{|\vec{r}|} |\vec{p}_B| .  \label{eqn:v4cfit40}
\end{eqnarray}

Above equations can be easily simplified to an equation with only
one unknown variable $|\vec{p}_B|$. With solved $|\vec{p}_B|$
value, all the other unknown variables can be calculated,
including four-momentum of the missing neutrino or other particle.
Thus, the SL $B$ decay is fully determined without tagging the $\bar{B}$
meson in the recoiling side. The distribution of $m_B$ in
Eq.~(\ref{eqn:eqn1}) can be used to extract the fractions of
signal and background components in experimental measurements.

\section{The SL decays of charged $B$ meson}

When the $B$ meson is charged, its momentum is no more along the
straight line between $IP$ and the $B$ decay vertex. The $B$ meson
deflects as a circle in $x$-$y$ plane, and moves uniformly as a
straight line in $z$-direction, resulting from the magnetic field
$\vec{B}$ in the detector. In this section, the algorithms are
developed for positive charged and negative charged $B$
mesons~\cite{footnote2} separately, considering different
direction of deflection in $x$-$y$ plane.

\subsection{The SL decays of $B^+$ meson}

\begin{figure*}
  \begin{center}
    \includegraphics[width=0.42\textwidth]{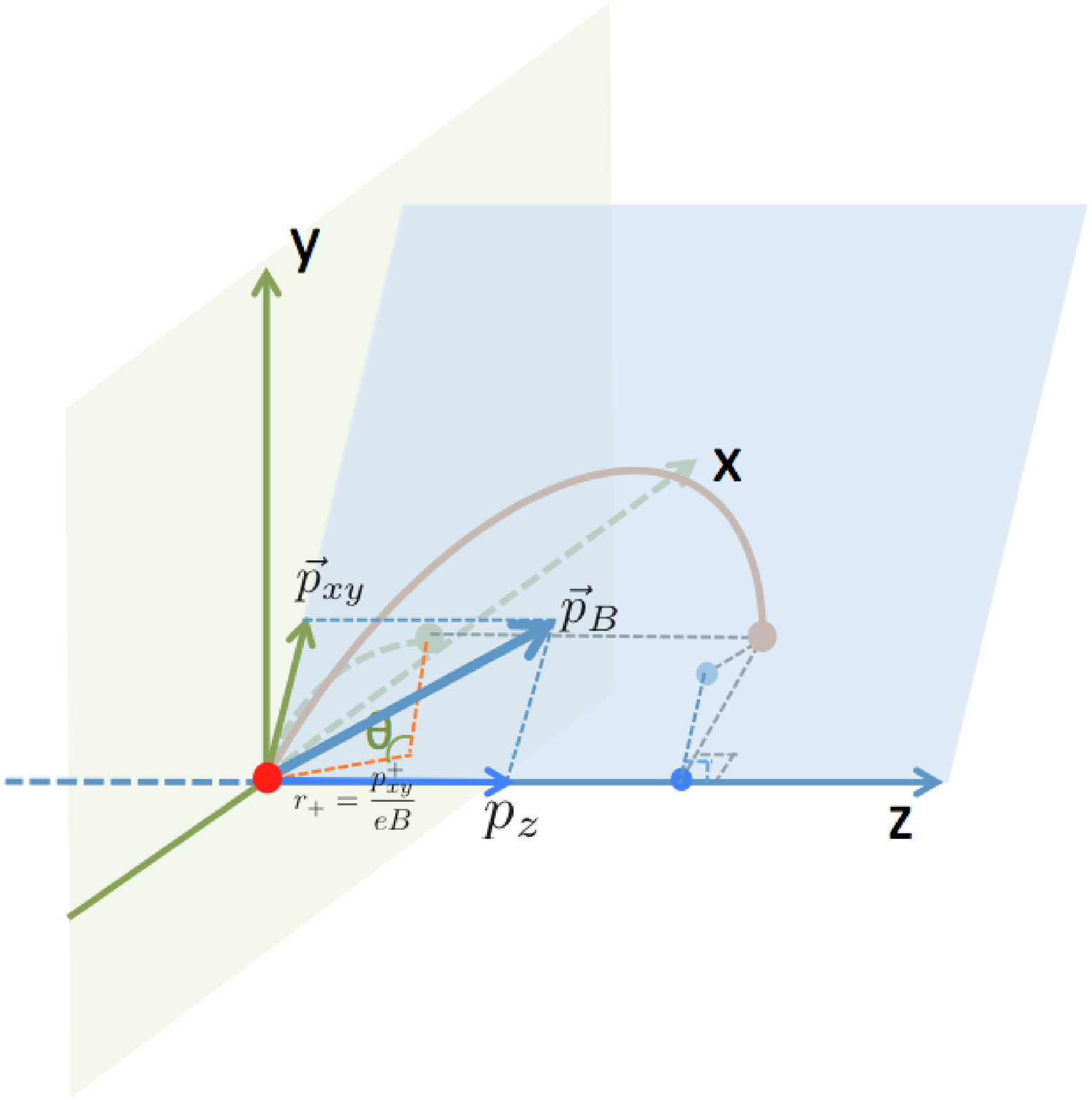}~
    \includegraphics[width=0.42\textwidth]{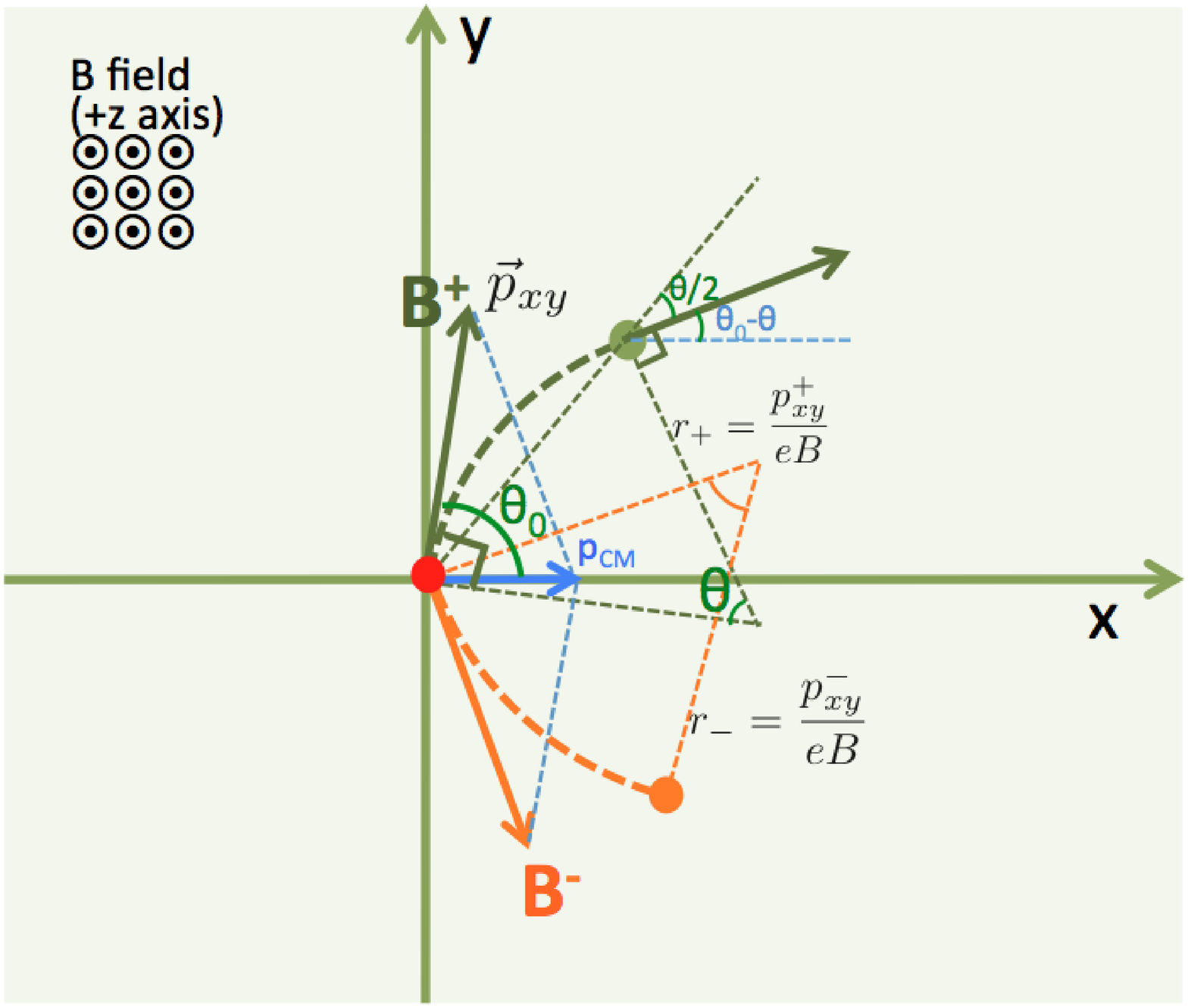}%
    \vskip-5pt
\caption{\label{fig:methodBp}(Color online) A schematic diagram of
SL decays of $B^+$ meson with vertices information.}
  \end{center}
\end{figure*}

For positive charged $B$ meson, it deflects on the right of the
flight direction in $x$-$y$ plane with radius
$r_{+}=|p_{xy}|/(eB)$, as shown in Fig.~\ref{fig:methodBp}.
Different from the neutral $B$ meson case, the direction of the
$B$ momentum can not be directly obtained by vector direction of
two vertices. In $x$-$y$ plane, we define $\theta_0$ as the angle between the
momentum of $B$ meson, $\vec{p}_{xy}$, and $+x$ axis; $\theta$ as the deflected angle due to magnetic field. 
Thus, the motion length of
arc in $x$-$y$ plane is $r_+\theta$. According to the motion
character in $x$-$y$ plane and in $z$-axis, the SL $B$ decay can still
be determined by following equations:
\begin{eqnarray}
f_{x}: &~~&  p_{hx} + p_{\ell x} + p_{\nu x}  = \sqrt{ p_{x}^2 + p_{y}^2} \cos(\theta_0 - \theta ),   \label{eqn:Bpv4cfit10}\\
f_{y}: &~~&  p_{hy} + p_{\ell y} + p_{\nu y} = \sqrt{ p_{x}^2 + p_{y}^2} \sin( \theta_0 - \theta ),  \label{eqn:Bpv4cfit20}\\
f_{z}: &~~& p_{hz} + p_{\ell z} + p_{\nu z} = p_z,   \label{eqn:Bpv4cfit30} \\
f_{e}: &~~& \sqrt{m_h^2 + |\vec{p}_h|^2} + \sqrt{m_{\ell}^2 + |\vec{p}_{\ell}|^2}  + \sqrt{m_{\nu}^2 + |\vec{p}_{\nu}|^2 }   = \sqrt{ m_B^2+|\vec{p}_B|^2},  \label{eqn:v4cfit40}\\
f_{r1}: &~~& r_{x} = \frac{2}{ eB }\sqrt{ p^2_{x} +  p^2_{y} } \cdot\sin\frac{\theta}{2}\cos(\theta_0 - \frac{\theta}{2}),  \label{eqn:Bpv4cfit60} \\
f_{r2}: &~~& r_{y} = \frac{2}{ eB }\sqrt{ p^2_{x} + p^2_{y} }  \cdot\sin\frac{\theta}{2}\sin(\theta_0 - \frac{\theta}{2}), \label{eqn:Bpv4cfit70} \\
f_{r3}: &~~& r_{z} = v_z\cdot \frac{r_{+} \theta }{ v_{xy}} = \frac{ p_{z} \theta }{ eB }, \label{eqn:Bpv4cfit80}  \\
f_{r4}: &~~& \theta_0 = \arctan \frac{ p_{y} }{ p_{x}  }, \label{eqn:Bpv4cfit90}  \\
f_{e^*}: &~~& \frac{\sqrt{s}}{2\gamma} =  \sqrt{m_B^2+|\vec{p}_B|^2} + (\beta_x p_x +
\beta_z  p_z) .    \label{eqn:v4cfit50}
\end{eqnarray}

To reduce the unknown variables, a set of simplified equations are
obtained as Eqs.~(\ref{eqn:Bpfit1}-\ref{eqn:Bpfit7}) with the unknown
momenta of $B$ meson and $\nu$, and the angle $\theta$:
\begin{eqnarray}
f_{\nu x}: &~~&  p_{\nu x} =  \frac{eB}{2}( r_x \cot\frac{ \theta }{2} + r_y) -(p_{hx} + p_{\ell x}), \label{eqn:Bpfit1}\\
f_{\nu y}: &~~&  p_{\nu y} = \frac{eB}{2}( -r_x + r_y \cot\frac{ \theta }{2} ) - (p_{hy} + p_{\ell y} ), \label{eqn:Bpfit2}\\
f_{\nu z}: &~~&  p_{\nu z} = eB \frac{ r_z }{ \theta } - (p_{hz} + p_{\ell z}), \label{eqn:Bpfit3} \\
f_{Bx}: &~~& p_{x} = \frac{eB}{2} (r_x  \cot\frac{ \theta }{2} - r_y ),  \label{eqn:Bpfit4} \\
f_{By}: &~~& p_{y} = \frac{eB}{2} (r_x + r_y \cot\frac{ \theta }{2} ),   \label{eqn:Bpfit5} \\
f_{Bz}: &~~& p_{z} = eB \frac{ r_z }{ \theta}, \label{eqn:Bpfit6}  \\
f_{e}: &~~& \sqrt{m_h^2 + |\vec{p}_h|^2} + \sqrt{m_{\ell}^2 + |\vec{p}_{\ell}|^2} + \sqrt{m_{\nu}^2 + |\vec{p}_{\nu}|^2 }  =  \frac{\sqrt{s}}{2\gamma} - (\beta_x p_x + \beta_z  p_z). \label{eqn:Bpfit7}
\end{eqnarray}
Then an equation with only one variable $\theta$ can be easily
obtained by Eq.~(\ref{eqn:Bpfit7}) with other
Eqs.~(\ref{eqn:Bpfit1}-\ref{eqn:Bpfit6}). With solved $\theta$ and
the nine equations, all the unknown variables can be determined.
Thus, SL $B^+$ decays can be fully determined, including the
momentum of the undetectable neutrino.

\subsection{The SL decays of $B^-$ meson}

\begin{figure*}
  \begin{center}
    \includegraphics[width=0.45\textwidth]{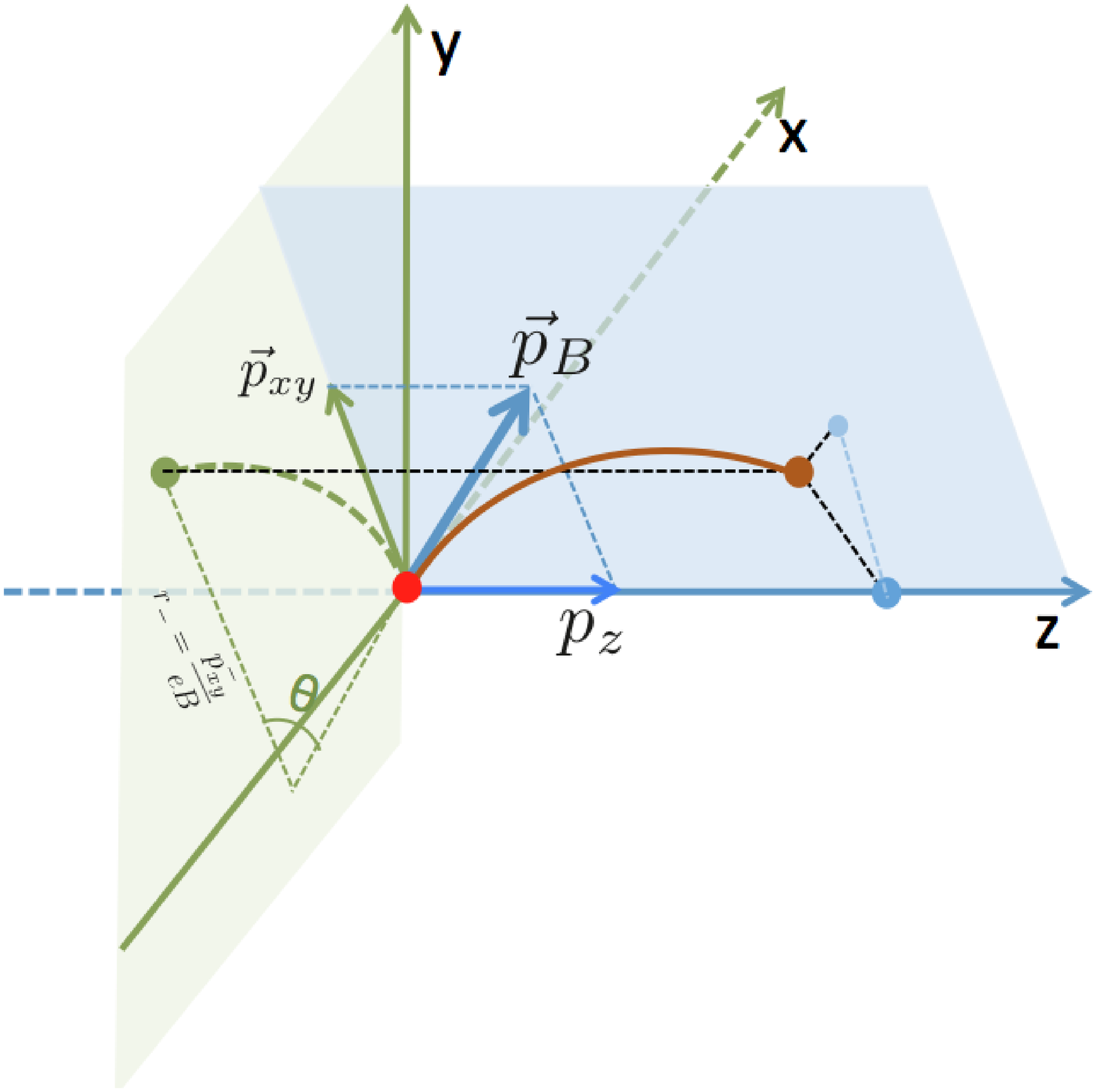}~
    \includegraphics[width=0.42\textwidth]{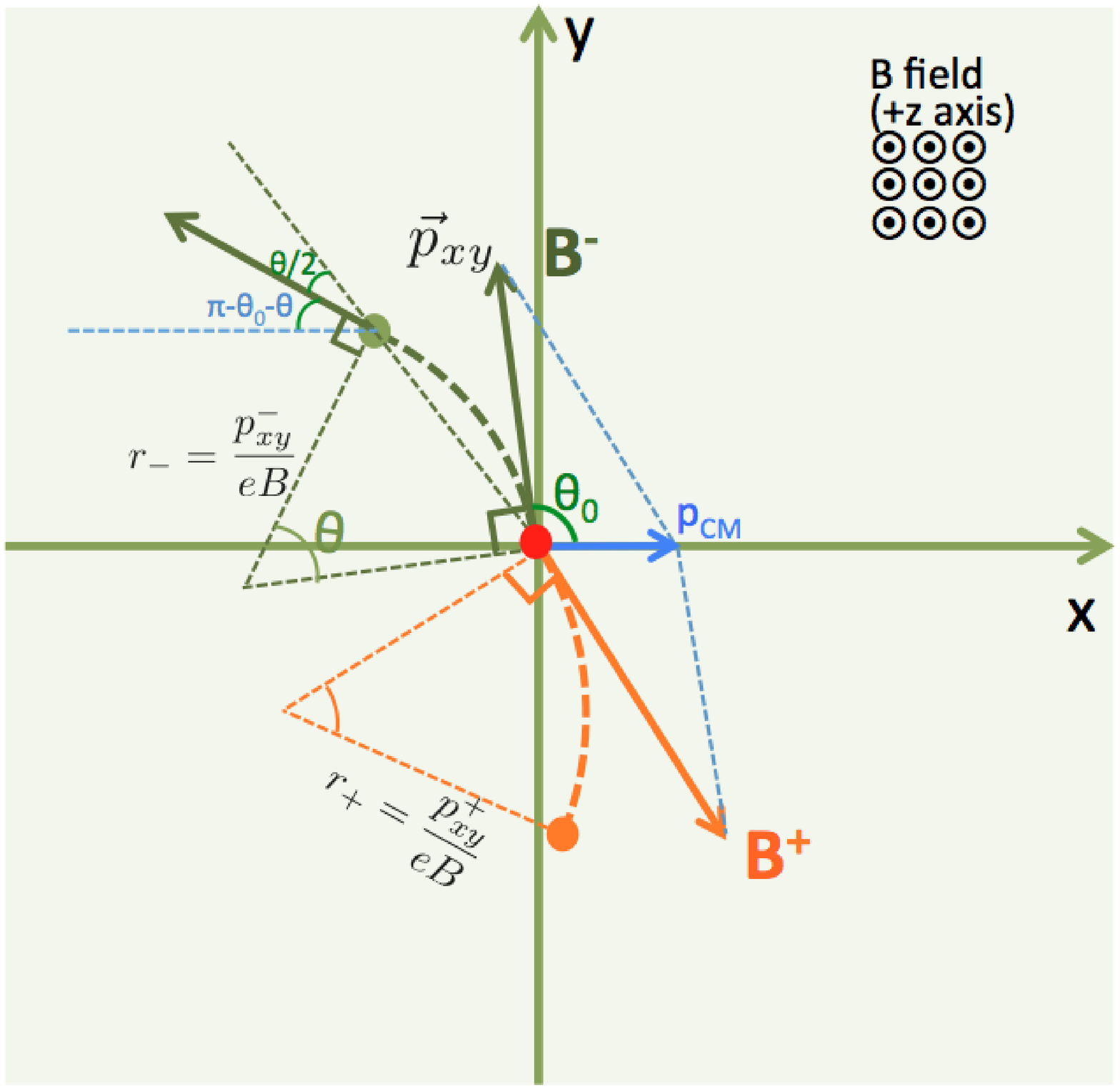}%
    \vskip-5pt
\caption{\label{fig:methodBm}(Color online) A schematic diagram of
SL decays of $B^-$ meson with vertices information.}
  \end{center}
\end{figure*}

For negative charged $B$ meson, it deflects on the left of the
flight direction in $x$-$y$ plane with radius
$r_{-}=|p_{xy}|/(eB)$, as shown in
Fig.~\ref{fig:methodBm}. Similar to the positive charged $B$ meson
case, following equations constrain the SL $B^-$ decays:
\begin{eqnarray}
f_{x}: &~~&  p_{hx} + p_{\ell x} + p_{\nu x}  = \sqrt{ p_{x}^2 + p_{y}^2} \cos( \theta_0 + \theta ),   \label{eqn:Bpv4cfit10}\\
f_{y}: &~~&  p_{hy} + p_{\ell y} + p_{\nu y}  = \sqrt{ p_{x}^2 + p_{y}^2} \sin( \theta_0 + \theta ),  \label{eqn:Bpv4cfit20}\\
f_{z}: &~~& p_{hz} + p_{\ell z} + p_{\nu z}  =  p_z,   \label{eqn:Bpv4cfit30} \\
f_{e}: &~~& \sqrt{m_{h}^2 + |\vec{p}_{h}|^2 } + \sqrt{m_{\ell}^2 + |\vec{p}_{\ell}|^2}  + \sqrt{m_{\nu}^2 + |\vec{p}_{\nu}|^2 }  = \sqrt{ m_B^2 + |\vec{p}_B|^2},  \label{eqn:v4cfit40}\\
f_{r1}: &~~& r_{x} = \frac{2}{ eB }\sqrt{ p^2_{x} + p^2_{y} }  \cdot \sin\frac{ \theta}{2}\cos(\theta_0 + \frac{ \theta }{2}), \label{eqn:Bpv4cfit60} \\
f_{r2}: &~~& r_{y} = \frac{2}{ eB }\sqrt{ p^2_{x} + p^2_{y} }  \cdot \sin\frac{ \theta}{2}\sin(\theta_0 + \frac{ \theta }{2}),  \label{eqn:Bpv4cfit70} \\
f_{r3}: &~~& r_{z} = v_z\cdot \frac{r_{-} \theta }{ v_{xy}} = \frac{ p_{z} \theta }{ eB }, \label{eqn:Bpv4cfit80}  \\
f_{r4}: &~~& \theta_0 = \arctan \frac{ p_{y} }{ p_{x}  }, \label{eqn:Bpv4cfit90}  \\
f_{e^*}: &~~&  \frac{\sqrt{s}}{2\gamma} = \sqrt{ m_B^2 + |\vec{p}_B|^2} + (\beta_x p_x + \beta_z  p_z) .  \label{eqn:v4cfit50}
\end{eqnarray}

Thus, to reduce unknown variables, a combination of equations is
simplified as follows:
\begin{eqnarray}
f_{\nu x}: &~~&  p_{\nu x} =   \frac{eB}{2} ( r_x \cot\frac{ \theta }{2} - r_y ) -(p_{hx} + p_{\ell x}),     \label{eqn:Bmfit1}\\
f_{\nu y}: &~~&  p_{\nu y} = \frac{eB}{2} (r_x + r_y \cot\frac{ \theta }{2}  ) -(p_{hy} + p_{\ell y} ),      \label{eqn:Bmfit2}\\
f_{\nu z}: &~~&  p_{\nu z} = eB \frac{ r_z }{ \theta } - ( p_{hz} + p_{\ell z} ),                    \label{eqn:Bpfit3} \\
f_{Bx}: &~~& p_{x} = \frac{eB}{2} (r_x \cot\frac{ \theta }{2} + r_y ),       \label{eqn:Bmfit4} \\
f_{By}: &~~& p_{y} =  \frac{eB}{2} (-r_x  + r_y \cot\frac{ \theta }{2}  ),       \label{eqn:Bpfit5} \\
f_{Bz}: &~~& p_{z} = eB \frac{ r_z }{ \theta } ,                 \label{eqn:Bmfit6}  \\
f_{e} : &~~& \sqrt{m_{h}^2 + |\vec{p}_{h}|^2 } + \sqrt{m_{\ell}^2 + |\vec{p}_{\ell}|^2}  + \sqrt{m_{\nu}^2 + |\vec{p}_{\nu}|^2 } =  \frac{\sqrt{s}}{2\gamma} - (\beta_x p_x + \beta_z  p_z). \label{eqn:Bmfit7}
\end{eqnarray}

Through solving an equation with one variable $\theta$ which can be
obtained from Eq.~(\ref{eqn:Bmfit7}) by using
Eqs.~(\ref{eqn:Bmfit1}-\ref{eqn:Bmfit6}), SL $B^-$ decays can be
fully determined, similar to the $B^+$ case.

\section{Conclusion and discussion}

We present a novel method to reconstruct SL $B$ decays with a missing
neutrino or hadronic $B$ decays with a missing particle without
tagging the $B$ meson in the recoiling side. Without suffering
from the very low tagging-efficiency (a few per mille to a few per
cent), our method is very promising to achieve a much larger utilization of $B$
decay sample for studying the SL $B$ decays.

This method depends strongly on the precision of the
reconstruction of the $IP$ and the $B$ decay vertex. Typical
resolution achieved at Belle experiment is $\sigma_x\sim
100~\mu$m, $\sigma_y\sim 1.9~\mu$m and $\sigma_z\sim 3.6$~mm for
$IP$ uncertainty and $\sim 100~\mu$m for vertex reconstruction
both along the beam direction and in transverse
plane~\cite{bib:PBF}. The resolution at Belle II is much improved
with nano-beam scheme for the $IP$ and better inner detectors for
secondary vertex reconstruction. The design resolution of $IP$ is
$\sigma_x\sim 10~\mu$m, $\sigma_y\sim 50$~nm and $\sigma_z\sim
0.15$ mm and excellent vertex resolution is $\sim 50~\mu$m at
Belle II~\cite{bib:BelleIITDR}.

Compared with the conventional way of studying SL $B$ decays, the
background level could be higher in the method presented here, so
some other techniques may need to be developed to further suppress
the background. An educated guess is even these background
suppressions further reduce the efficiency by an order of
magnitude, the efficiency of the new method is still higher than
the tag method by at least an order of magnitude. Of course, the
factors are mode dependent, detailed results can be obtained with
Monte Carlo study or with the real data at Belle and BaBar
experiments.

As we have mentioned, this new method can be used for the SL $B^0$
decays, such as $B^0\to\pi^{-}\ell^{+}\nu_{\ell}$ with an
undetectable neutrino for improved measurement of form factor and
$|V_{ub}|$~\cite{bib:BToSLBelleFR}; hadronic decays such as
$B^0\to D^- p\bar{n}$, which has not been observed yet but a 
similar decay $B^0\to D^{*-}p\bar{n}$ has been observed~\cite{bib:B0ToDstpnbar}, with a missing
anti-neutron for searching of exited charmed baryon
$\Lambda_c^{*}$, pentaquark $\Theta_c[\bar{c}uudd]$, or hadronic
structures in $D^-p$ system; SL decays of charged $B$ such as
$B^+\to p\bar{p}\ell^+\nu_{\ell}$ to target for the first
observation, which is now only an evidence from
Belle~\cite{bib:BmToppSL}.

Besides $B$ decays, this method can also be used in analyzing SL
$D$ decays or $\tau$ decays at BaBar, Belle, and Belle II
experiments where large momentum $D$ and $\tau$ may have even
longer decay length thus better momentum resolution. This method
also allows a study of the semileptonic decays of hyperons
($\Lambda$, $\Sigma$, $\Xi$, and so on) pair produced copiously in
$\tau$-charm factories such as BESIII experiment~\cite{bes3}, and
super $\tau$-charm factories under
discussion~\cite{hiepa,russian}.

\acknowledgments

This work is supported in part by National Natural Science
Foundation of China (NSFC) under contract Nos. 11475187 and
11521505; Key Research Program of Frontier Sciences, CAS, Grant
No. QYZDJ-SSW-SLH011; and the CAS Center for Excellence in
Particle Physics (CCEPP).



\begin{thebibliography}{999}
\bibitem{footnote1} 
In this article, only two charged-particle case
is presented as an example, the method can be extended easily for
the SL $B$ decays with more than two charged particles.
\bibitem{footnote2} Due to charge conservation, at least three charged
particles are required in final state particles of SL $B^{\pm}$
decays for our method. We still use $h$ for charged particles
other than $\ell^{\pm}$, this will not effect our analysis but
make the discussion more convenient.
\end{thebibliography}
\end{document}